%% file: template.tex
\documentclass[preprint,journal]{vgtc}       % preprint (journal style)

\ifpdf%                                % if we use pdflatex
  \pdfoutput=1\relax                   % create PDFs from pdfLaTeX
  \usepackage{graphicx}                % allow us to embed graphics files
  \DeclareGraphicsExtensions{.pdf,.png,.jpg,.jpeg} % for pdflatex we expect .pdf, .png, or .jpg files
\else%                                 % else we use pure latex
  \usepackage{graphicx}                % allow us to embed graphics files
  \DeclareGraphicsExtensions{.eps}     % for pure latex we expect eps files
\fi%

\graphicspath{{figures/}{pictures/}{images/}{./}} % where to search for the images

\usepackage{microtype}                 % use micro-typography (slightly more compact, better to read)
\PassOptionsToPackage{warn}{textcomp}  % to address font issues with \textrightarrow
\usepackage{textcomp}                  % use better special symbols
\usepackage{mathptmx}                  % use matching math font
\usepackage{times}                     % we use Times as the main font
\usepackage{cite}                      % needed to automatically sort the references
\usepackage{tabu}                      % only used for the table example
\usepackage{booktabs}                  % only used for the table example
\usepackage{caption}
\usepackage{amsmath,amssymb,amsthm}
\usepackage{mathtools}
\usepackage{subfig}
\usepackage{float}
\usepackage{algorithm}
\usepackage[noend]{algpseudocode}
\usepackage{tabularx}
\usepackage{epstopdf}
\usepackage{multirow}
\usepackage{enumitem}
\usepackage[usenames, dvipsnames]{color}
\usepackage{mathtools}
\usepackage{gensymb}
\usepackage{url}

\definecolor{mygreen}{rgb}{0.0, 0.8, 0.0}
\definecolor{applegreen}{rgb}{0.55, 0.71, 0.0}
\definecolor{brightgreen}{rgb}{0.4, 1.0, 0.0}
\definecolor{myblue}{rgb}{0.0, 0.0, 1.0}

\newcommand{\clrr}{\textcolor{black}}
\newcommand{\clrb}{\textcolor{black}}
\newcommand{\clrg}{\textcolor{black}}

\usepackage{xcolor}
\definecolor{light-blue}{HTML}{99e6ff}

\definecolor{light-red}{HTML}{ff9980}

\definecolor{light-green}{HTML}{8cff66}

\definecolor{light-gray}{HTML}{bfbfbf}

\ieeedoi{10.1109/TVCG.2019.2934591}
\onlineid{1223}

\vgtccategory{Research}
\vgtcpapertype{system}

\title{NNVA: Neural Network Assisted Visual Analysis of Yeast Cell Polarization Simulation}

\author{Subhashis Hazarika, Haoyu Li, Ko-Chih Wang, Han-Wei Shen, and Ching-Shan Chou}
\authorfooter{
\item
 Subhashis Hazarika, Haoyu Li, Ko-Chih Wang, Han-Wei Shen are with Department of Computer Science, Ohio State University. E-mail: hazarika.3, li.8460, wang.3182, shen.94@osu.edu.
\item
 Ching-Shan Chou is with Department of Mathematics, Ohio State University. E-mail: chou@math.osu.edu
}

\shortauthortitle{Biv \MakeLowercase{\textit{et al.}}: Global Illumination for Fun and Profit}
\abstract{ Complex computational models are often designed to simulate real-world physical phenomena in many scientific disciplines. However, these simulation models tend to be computationally very expensive and involve a large number of simulation input parameters, which need to be analyzed and properly calibrated before the models can be applied for real scientific studies. We propose a visual analysis system to facilitate interactive exploratory analysis of high-dimensional input parameter space for a complex yeast cell polarization simulation. The proposed system can assist the computational biologists, who designed the simulation model, to visually calibrate the input parameters by modifying the parameter values and immediately visualizing the predicted simulation outcome without having the need to run the original expensive simulation for every instance. Our proposed visual analysis system is driven by a trained neural network-based surrogate model as the backend analysis framework. In this work, we demonstrate the advantage of using neural networks as surrogate models for visual analysis by incorporating some of the recent advances in the field of uncertainty quantification, interpretability and explainability of neural network-based models. We utilize the trained network to perform interactive parameter sensitivity analysis of the original simulation as well as recommend optimal parameter configurations using the activation maximization framework of neural networks. We also facilitate detail analysis of the trained network to extract useful insights about the simulation model, learned by the network, during the training process. \clrb{We performed two case studies, and discovered multiple new parameter configurations, which can trigger high cell polarization results in the original simulation model. We evaluated our results by comparing with the original simulation model outcomes as well as the findings from previous parameter analysis performed by our experts.}

} % end of abstract


\CCScatlist{ % not used in journal version
 \CCScat{K.6.1}{Management of Computing and Information Systems}%
{Project and People Management}{Life Cycle};
 \CCScat{K.7.m}{The Computing Profession}{Miscellaneous}{Ethics}
}

\teaser{
  \centering
  \includegraphics[width=0.9\linewidth]{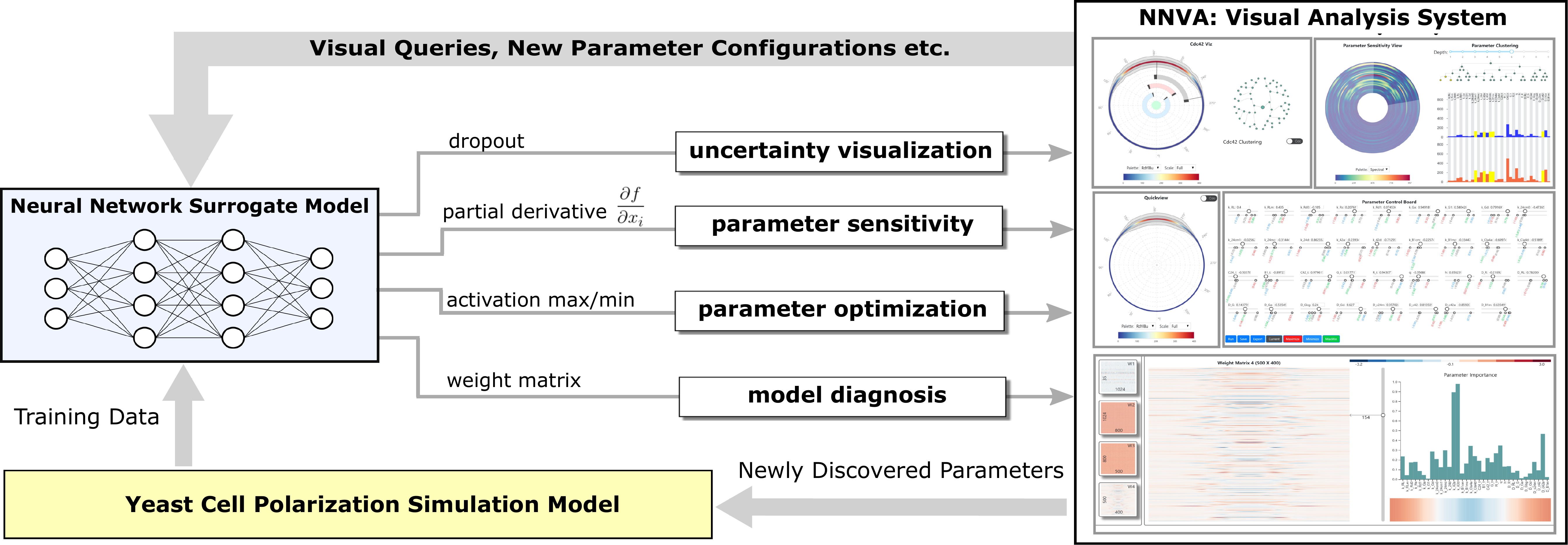}
  \caption{Approach Overview: A trained neural network-based surrogate model acts as the backend analysis framework, driving our interactive visual analysis system for analyzing a computationally expensive yeast simulation model.}
	\label{fig:teaser}
}

\vgtcinsertpkg

\begin{document}


\firstsection{Introduction}

\maketitle
\input{tex/1_introduction}

\section{Related Work}
\input{tex/2_relatedwork}
\section{Simulation Model Background}
\input{tex/3_background}

\section{Requirement Analysis and Approach Overview}
\input{tex/4_requirement}

%
\section{Neural Network-based Surrogate Model}
\input{tex/5_NNSM}

\section{Neural Network Assisted Visual Analysis }
\input{tex/6_NNVA}

%
\section{Case Study and Evaluation}
\input{tex/7_case_study}

\section{Domain Expert Feedback}
\input{tex/8_feedback}

\section{Discussion}
\input{tex/9_discussion}
%
\section{Conclusion and Future Work}
\input{tex/10_conclusion}

%% \section{Introduction} %for journal use above \firstsection{..} instead
%This template is for papers of VGTC-sponsored conferences such as IEEE VIS, IEEE VR, and ISMAR which are published as special issues of TVCG. The template does not contain the respective dates of the conference/journal issue, these will be entered by IEEE as part of the publication production process. Therefore, \textbf{please leave the copyright statement at the bottom-left of this first page untouched}.

\acknowledgments{
This work was supported in part by US Department of Energy Los Alamos National Laboratory contract 47145 and UT-Battelle LLC contract 4000159447 program manager Laura Biven. We would also like to thank Marissa Renardy and Luke Andrejek for their help and feedback.}

\bibliographystyle{abbrv}

\bibliography{template}
\end{document}

%% file: tex/1_introduction.tex
\setlength{\belowcaptionskip}{-10pt}
In the field of computational biology, scientists often design mathematical simulation models to offer quantitative descriptions of complex biological processes. These simulations are subsequently used to perform in-depth analyses of the real biological phenomenon. However, designing an optimal simulation model can be challenging. Scientists need to have a clear picture of how the different simulation input parameters are affecting the simulation output. For compute-intensive simulation models with high-dimensional input and output spaces, this can become a computationally prohibitive and non-trivial analysis task.

We collaborated with computational biologists to design an interactive visual analysis framework\clrr{,} which can assist them in analyzing and visualizing a complex \textit{yeast cell polarization} simulation model. The model simulates the concentration of important protein molecules along the membrane of a yeast cell (single-cell microorganism) during its mating process. Cell polarization refers to asymmetric localization of protein concentration in a small region of the cell membrane and is a fundamental stage in the life-cycle of many microorganisms. Our experts are interested in exploring and analyzing the simulation input parameters\clrr{,} which can simulate varying levels of cell polarization results, particularly, the ones with high polarization. However, there are 35 different unknown/uncalibrated input parameters for the simulation. Besides the high-dimensional nature of the problem, the simulation model itself is computationally expensive. It takes hours on a supercomputing cluster to complete a single execution of the model. This seriously hampers the possibility of performing any exploratory analysis task that requires frequent execution of the simulation on new and unseen parameter configurations to study its properties in detail. In the field of simulation sciences, a popular and effective strategy to address this issue has been to create a simpler statistical/mathematical \textit{surrogate model}, mimicking the original expensive simulation model~\cite{surrogate1, surrogate2, surrogate3, surrogate4, surrogate5}. The surrogate is then utilized to perform detailed analysis tasks instead of the expensive simulation model. A well-trained surrogate model can greatly facilitate the analysis workflow of complex simulation models.

Compared to popular surrogate model options like \textit{polynomial fitting} or \textit{Gaussian Processes}, \textit{neural networks} are particularly well-suited for designing interactive visual analysis systems. This is primarily because, besides accurately predicting the output of high-dimensional non-linear functions, they can also be utilized to extract and analyze interesting properties about the original simulation by opening up the \textit{black-box} of the trained neural networks. Recent advances in the field of \textit{interpretability} and \textit{explainability} of neural network-based models~\cite{interpretability} have resulted in many useful post-hoc analysis techniques, making them more transparent in the process. This has led to a surge in their usage as proper analysis tools in many application domains~\cite{Application1,Application2,Application3,Application4}. In this work, we propose a \textit{\underline{n}eural \underline{n}etwork}-assisted \textit{\underline{v}isual \underline{a}nalysis} system (NNVA)\clrr{,} which utilizes a \textit{neural network-based surrogate model} to perform exploratory analysis and visualization of the aforementioned yeast simulation model. The surrogate model acts as the backend analysis framework, facilitating various visual interactions and analysis activities in the system. Of late, there is a growing interest in the usage of different neural network-based models to solve complex real-world problems. Our work, therefore, exemplifies how to build an interactive visual analysis system around these powerful models, i.e, \clrb{\textit{Machine Learning for Visual Analytics (ML4VA)}}.

Our proposed system facilitates interactive exploratory analysis by allowing the experts to modify the input parameter values and immediately visualize the predicted simulation outcome. This helps them discover new parameter configurations without having the need to execute the original expensive simulation for every instance. \clrb{Using different interactive selection brushes, experts can perform parameter sensitivity analysis at multiple levels of detail as well as get optimal parameter recommendations to produce desired simulation outcomes for selected regions of the membrane. To establish the trustworthiness of any visual analysis system, it is important to convey the underlying uncertainty associated with the predictions of the surrogate model. We utilize a recently proposed uncertainty quantification technique for neural networks using dropout layers~\cite{dropout2016} to incorporate uncertainty visualization in our system.} We also allow the experts to validate the surrogate model itself, by analyzing the various weight matrices and extracting the knowledge learned by the surrogate during the training process. We performed extensive evaluations of the proposed framework by comparing with the original simulation model and the results of a previous analysis effort which used polynomial surrogate models~\cite{renardy2018}.

To summarize, the major contributions of our work are as follows:
\begin{itemize}
	\setlength\itemsep{0.1em}
	\item We demonstrate how a trained neural network can act as an analysis backend to drive an interactive visual analysis system.
	%\item We incorporated many recent contributions from the field of interpretability, explainability and uncertainty quantification of neural network models in our visual analysis framework.
	%\item We offered different levels of parameter sensitivity analysis using the knowledge learned by the hidden layers of the neural network in the training process.
	\item We discovered multiple previously unknown parameter configurations, which generated strong cell polarization results in the original simulation model.

	\item \clrb{We provide easy integration of our visual analysis workflow with the simulation modeling workflow of the experts by allowing them to store the discovered configurations in a file format, which can be used to directly execute the original simulation model.}
\end{itemize}

%% file: tex/2_relatedwork.tex
In this section, we focus on previous research works from the field of visual analysis, which are relevant to our proposed approach.

\textbf{Visual Exploration of Parameter Space:} Over the years, multiple visual analysis systems have been proposed to facilitate interactive visual exploration of the input parameter space for simulation models. Each of them are application specific, and caters to the requirements of their domain experts. Orban et al.~\cite{orban2019drag} projected input parameters and output data in material science to 2D spaces, and allowed users to manipulate in the input space and observe the change in the output space. Wang et al.~\cite{wang2017multi} developed a nested parallel coordinate plot for parameter analysis of multi-resolution climate ensemble datasets. Biswas et al.~\cite{biswas2018interactive} used a \clrr{Gaussian Process}-based surrogate model to perform interactive exploration in a shock physic application. Coffey et al.~\cite{coffey2013design} designed an interface which uses a mapping between model features and simulation inputs to enable direct simulation input parameter manipulations. Piringer et al.~\cite{piringer2010hypermoval} proposed an approach called HyperMoVal, which can evaluate the bad fit of surrogate models and provide visual validation for their physical plausibility. Berger et al.~\cite{berger2011uncertainty} proposed an uncertainty-aware statistical approach to predict results of given parameters for real-time analysis. 

Sedlmair et al.~\cite{sedlmair2014visual} provided an extensive survey and proposed a conceptual framework to categorize the various analysis tasks and navigation strategies used in such visual analysis systems. Our proposed visual analysis system encompasses three of the six analysis tasks formalized by Sedlmair et al., namely, \textit{optimization}, \textit{uncertainty}, and \textit{sensitivity}. Among the four navigation strategies that they identified, our system covers two of them, namely \textit{informed trial and error} and \textit{local-to-global}.  Simulation parameter analysis is also a popular topic in scientific visualization community. Parameter sensitivity analysis techniques~\cite{murphy2004quantification, kleijnen1992techniques, elsey2011large} have been widely used to perform various uncertainty-aware scientific analysis and visualization~\cite{biswas2017visualization, hamby1994review, sens1}. The recent survey on visualization techniques for ensemble simulation data by Wang et al. \cite{wang2018visualization} also covers the sub-category of simulation parameter analysis \clrr{in the visualization} community.

\clrb{\textbf{Machine Learning Models for Visual Analysis:} Visualization community often uses machine learning techniques to enhance their visual analytic tools~\cite{ml4va}. Machine learning models act as a medium to extract interesting insights about the data, which is then presented to the end users through interactive visual analytic systems. Besides enhancing the data-analysis experience, this acts as a platform for users without much machine learning background to reap the benefits of sophisticated machine learning models. Among the recent neural network-based models, CNN (Convolutional Neural Network)~\cite{xie2018semantic} and Word2Vec~\cite{zhou2019visual,lu2019inkplanner} models have been used to create interactive visual analysis systems for different application domains. Moreover, traditional models like SVM (Support Vector Machine)~\cite{xie2019visual}, LDA (Latent Dirichlet Allocation)~\cite{lee2012ivisclustering,liu2012tiara}, KNN (K Nearest Neighbor)~\cite{marai2018precision}, Bayes' rule~\cite{goodall2019situ}, learning-from-crowds model~\cite{liu2019interactive}, and online metric learning~\cite{liang2018photorecomposer} have also been extensively utilized by visualization researchers to enhance the data-analysis experience in their systems. Along similar lines, our proposed system utilizes a trained multilayer perceptron model to design an interactive visual analysis framework for a scientific application.}

\textbf{Visual Analysis for Machine Learning Models:} Since the past few years, the visualization community has played a significant role in explaining the inner workings of complicated machine learning models. Multiple visual analytic tools have been developed to visualize different machine learning algorithms, such as Adaboost, SVM, decision \clrr{tree, and} random forest~\cite{jakulin2005nomograms, hentschel2015image, cortez2013using, safavian1991survey, vondrick2013hoggles}. Recently, Hohman et al.~\cite{Hohman2018VisualAI} published a comprehensive survey on the various visual analytic approaches to explain deep neural network models, which are gaining significant popularity in the machine learning community. At a high-level, we \clrr{can divide} these approaches into three categories. The goal of one category of visualization tools is to open the black box by interpreting the trained model~\cite{wongsuphasawat2018visualizing, liu2017towards, ming2017understanding, wang2018ganviz, strobelt2018lstmvis, kahng2019gan}. While, another category of visualization tools not only interpret, but also diagnose the trained model~\cite{kahng2018cti, zhang2019manifold, strobelt2019s, wang2019deepvid}. Recently, a third category of visualization tools, focusing more on assisting the machine learning experts to improve their models is gaining popularity\cite{wang2019dqnviz, bilal2018convolutional}.

Besides the visualization community, the machine learning community is also working in parallel to create various post-hoc analysis techniques to interpret and explain complicated models. Unlike most of the visualization tools, these post-hoc analysis functions are intended to be more generic and applicable for different network architectures. Montavon et al.~\cite{interpretability} covers in great details the various post-hoc analysis techniques that can be performed on trained neural networks to make them more interpretable and explainable. These analysis techniques for trained neural networks have seen wide-spread application in scientific domains, ranging from cancer diagnosis to quantum physics~\cite{Application1,Application2,Application3,Application4}. In our proposed visual analysis system, we use different post-hoc analysis functions on the trained neural network-based surrogate model to study and analyze the original yeast simulation. We also visualize the network structure (weight matrices) to extract and validate the knowledge learned by the surrogate model during the training process.   

%% file: tex/3_background.tex
\begin{figure}[t!]
\centering
        \includegraphics[width = 0.85\linewidth]{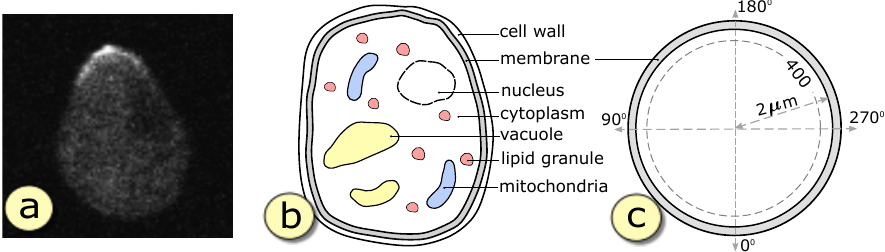} 
\caption{(a) Microscopic image of a highly polarization yeast cell. (b) Pedagogical illustration of the yeast cell structure. (c) The computational domain used in the simulation to model the cell membrane. }
\label{yeast_background}
\end{figure}

Traditional laboratory-based approach for studying yeast cell polarization consists of a laborious workflow. The cells are first cultured/grown in a growth media and then treated with various straining agents. They are then visualized using high-resolution microscopes. Fig.~\ref{yeast_background}(a) shows the microscopic image of a highly polarized yeast cell. A mathematical simulation model, therefore, can significantly accelerate the study of such biological phenomena. 

To capture the spatio-temporal dynamics of yeast cell polarization during the mating cycle, scientists created a mechanistic spatial model to simulate the concentration of important protein species along the cell membrane. In this work, we focus only on one important protein species called \textit{Cdc42}. Protein concentration is measured as the number of molecules per unit area of the membrane. The model simulates the cell membrane as a circle, centered at the origin with radius $2 \mu m$. Fig.~\ref{yeast_background}(b) shows a pedagogical diagram of the yeast cell structure with the peripheral cell membrane, while, Fig.~\ref{yeast_background}(c) shows the corresponding computational domain of the simulation model used by the scientists. The computational domain (circle) \clrr{has a spatial} resolution of 400, parameterized by angles in the range $[0 \degree,360 \degree]$. Scientists are interested in simulating results with high degree of polarization, in particular of \textit{Cdc42} protein. To quantify the extent of Cdc42 polarization, they constructed a scalar function of active Cdc42 (C42a) values called \textit{polarization factor} (PF), denoted as;
\begin{equation}
PF = \Big( 1 - 2\frac{S_p(C42a)}{SA}\Big)\times \frac{(ax)^5}{1 + (ax)^5}
\label{eq1}
\end{equation} 
where, $SA$ is the surface area of the membrane simulated by the model, $S_p(C42a)$ is the surface area at the front of the cell that encompasses half of the polarized component C42a, $x$ is the maximum C42a concentration value and $a$ is an experiment constant dependent on simulation parameter. An unpolarized cell would have a PF value of 0 and an infinitely polarized cell would have a PF value of 1.

However, the simulation model comprises of 35 different input parameters. For the model to be useful to study the biological process of yeast cell polarization, scientists need to have a clear understanding of how the simulation input parameters effect the simulation results. More specifically, they want to figure out the parameter configurations which can generate high cell polarization results in the model. Studying this high-dimensional parameter space is not trivial, especially when the individual simulation execution itself takes few hours to execute. Any analysis task that requires frequent execution of the model on new parameter configurations is effected by the long execution time of the simulation model.

% \begin{figure}[t!]
% \centering
% \subfloat[]{
%         \includegraphics[width = 0.30\linewidth]{images/simulation_model/yeast_image.png} }
% \subfloat[]{
%         \includegraphics[width = 0.35\linewidth]{images/simulation_model/yeast.pdf} }
% \subfloat[]{
%         \includegraphics[width = 0.34\linewidth]{images/simulation_model/domain.pdf} }
% \caption{Yeast Cell: (a) Image from a high-resolution microscope showing the polarization of Cdc42 protein in a yeast cell. The pixel intensity reflects the protein concentration at a location. (b) A pedagogical diagram of the yeast cell structure. The region beneath the cell wall corresponds to the cell membrane. Scientists are interested in simulating the concentration of protein in this region. (c) The corresponding computational domain of the simulation model representing the cell membrane as a circle of radius  $2 \mu m$. }
% \label{yeast_background}
% \end{figure}

%Yeast Cell: (a) Image from a high-resolution microscope showing the polarization of Cdc42 protein in a yeast cell. The pixel intensity reflects the protein concentration at a location. (b) A pedagogical diagram of the yeast cell structure. The region beneath the cell wall corresponds to the cell membrane. Scientists are interested in simulating the concentration of protein in this region. (c) The corresponding computational domain of the simulation model representing the cell membrane as a circle of radius  $2 \mu m$. 

\textbf{Previous Simulation Model Analysis:} Previous efforts into analyzing this simulation model involved creating a polynomial surrogate model~\cite{renardy2018}. The surrogate model was created by uniformly sampling the parameter space and fitting a polynomial function to the polarization factor (PF) values (Equation~\ref{eq1}). The surrogate model facilitated in analyzing the parameter sensitivity of the model and helped estimate parameter configurations using a Markov Chain Monte-Carlo (MCMC) approach. However, the approach was not able to identify satisfactory parameter configurations with which the simulation can generate significantly high polarization results. Moreover, given a parameter configuration, the polynomial surrogate model only predicts the final PF value and not the Cdc42 protein concentration values across the membrane, which is the final output of the simulation. As a result, parameter sensitivity analysis performed with the polynomial surrogate model was not able to study the influence of the parameters on different regions of the membrane in finer details. Such analysis is important to get a better understanding of how the simulation parameters actually affect the protein concentration across the membrane and not just the final PF value. In this work, we train a neural network-based surrogate model to predict Cdc42 concentration values for the 400 spatial locations across the membrane as modeled by the computational domain (\clrr{Fig.~\ref{yeast_background}(c)}). We use this neural network-based surrogate model to create a visual analysis system for the simulation model. We utilize some of the important findings in the previous work~\cite{renardy2018} to extensively evaluate and validate the results produced by our proposed system.

%\clrb{In this work, assisted by a neural network-based surrogate model, we propose a visual analysis framework to facilitate interactive visual analysis and diagnosis of the yeast simulation model. Instead of predicting the PF values~\cite{renardy2018}, our neural network-based surrogate model predicts the Cdc42 concentration values across the 400 spatial locations of the membrane. As a result, our visual analysis approach is able to perform analysis at different levels-of-detail than the previous surrogate model based analysis efforts. We utilize some of the important findings in the previous work~\cite{renardy2018} to extensively evaluate and validate the results produced by our proposed method.}  

%In this work, we propose a neural network-based surrogate model to predict the Cdc42 concentration values across the 400 spatial locations of the membrane. Such neural network-based surrogate model can also drive a visual analysis framework, facilitating interactive analysis of the simulation model at various levels of detail. We utilize some of the important findings in the previous work~\cite{renardy2018} to extensively evaluate and validate the results produced by our proposed method.

% \begin{figure*}[t!]
% \centering
%         \includegraphics[width = 0.85\linewidth]{images/overview.pdf} 
% \caption{Overview}
% \label{overview}
% \end{figure*}  

%The simulation model comprises of a collection partial differential equations modeling the surface diffusion on the cell membrane and reactions with other proteins in the system.

%% file: tex/4_requirement.tex
\textbf{Requirements:} Throughout the course of this project, we had multiple interactions with the scientists from computational biology to understand the various aspects of their simulation model and get a clear picture of their needs and requirements from a visual analysis system. Based on these discussions, the most important requirements are as follows:
\begin{itemize}
	\setlength\itemsep{0.15em}
	\item[\textbf{R1}] Discover new parameter configurations \clrr{which can generate} high Cdc42 polarization results in the simulation model. The system should have the ability to visually guide the users in the process of finding desired parameter configurations for the simulation. 

	\item[\textbf{R2}] Ability to get a quick preview of the predicted simulation output \clrr{for} particular parameter configuration to facilitate model calibration. This helps them decide whether to execute the expensive simulation model with certain parameter configurations or not.

	\item[\textbf{R3}] Perform \clrr{detailed sensitivity} analysis of the input parameters with respect to Cdc42 concentration for different regions of the cell membrane.

	\item[\textbf{R4}] Analyze the distribution of protein concentration values across the computational domain of the model. This is required to decide on an ideal partitioning scheme for the computational domain. 

	\item[\textbf{R5}] Ability to extract and validate the knowledge learned by the surrogate during its training process. This is required to make sure that the trained network is not making any random predictions.
\end{itemize}

\textbf{Overview:} Based on these requirements, we have proposed an interactive visual analysis system, which is driven by a neural network-based surrogate model. We first train a fully connected neural network on a finite set of training data, obtained by running the simulation on random parameter configurations. The neural network learns to predict the concentration of the Cdc42 protein along the cell membrane for a given input parameter configurations. We then design a visual analysis framework which allows the users to visually query for different properties about the simulation model to address the aforementioned user requirements. These queries are executed in the backend by the trained surrogate model \clrr{to provide} prompt feedback via the visual interface. The system visually guides the users to discover new parameter configurations, which can be later used to execute the original simulation model. Fig.~\ref{fig:teaser} shows the high-level overview of the proposed system.

%The visual analysis system allows scientists to study the sensitivity of an input parameter configuration in great details across the spatial resolution of the membrane. We also guide the scientists towards discovering new parameter configuration by recommending optimal parameter configurations that maximizes/minimizes protein concentration for user selected regions of interest. Scientists can easily play around with different parameter configurations and see the predicted Cdc42 profile before running the actual simulation model. To assist them in their decision making process, the uncertainty associated with the predicted results is also conveyed in the same view. The newly discovered parameter configurations can be directly saved into configurations files which can be used to execute the original simulation, thus, offering a smooth integration of analysis workflow with the simulation process. Figure~\ref{fig:teaser} shows the high-level overview of the proposed system.

%% file: tex/5_NNSM.tex
% \begin{figure}[t!]
% \centering
%         \includegraphics[width = 0.65\linewidth]{images/NNSM/architecture.pdf} 
% \caption{Architecture of the neural network-based surrogate model.}
% \label{network_structure}
% \end{figure}

% \begin{figure}[t!]
% \centering
% \subfloat[]{
%         \includegraphics[width = 0.55\linewidth]{images/NNSM/architecture.pdf} }
% \subfloat[]{
%         \includegraphics[width = 0.45\linewidth]{images/NNSM/dropout_example.pdf} }
% \caption{}
% \label{PFdistribution}
% \end{figure}

Surrogate models are widely used in many areas of engineering and simulation science as cost effective alternatives to expensive simulation models for various analyses~\cite{surrogate1, surrogate2, surrogate3, surrogate4, surrogate5}. Also known as response surface models or emulators, they mimic the behavior of the actual simulation model as closely as possible while being computationally easier to evaluate. The surrogate model proposed in our work is a \textit{multi-layer \clrr{fully-connected} feed-forward regression neural network}. In this section, we first discuss in details the network structure and the training process of our surrogate model and then elaborate on the various post-hoc analysis techniques that can be performed on trained networks to facilitate the eventual visual analysis system.

%A well designed surrogate model can greatly facilitate exploratory analysis activities which require repeated execution of the expensive simulation model. 

%The network takes in 35 different simulation input parameters and predicts the corresponding Cdc42 protein concentration along the 400 uniformly distributed spatial locations of the cell membrane modeled by the original simulation. In this section, we explain in details the overall network architecture and the corresponding training process of the surrogate model. We also elaborate on the various analysis functions that we perform on a trained neural network to facilitate the eventual visual analysis system. In particular, we discuss uncertainty quantification in neural networks using a dropout bayesian approach, parameter sensitivity analysis and parameter optimization in a trained neural network.

% \begin{figure}[t!]
% \centering
% \subfloat[First 2000 training data]{
%         \includegraphics[width = 0.47\linewidth]{images/NNSM/training_data_1.pdf} }
% \subfloat[Complete 3000 training data]{
%         \includegraphics[width = 0.47\linewidth]{images/NNSM/training_data_2.pdf} }
% \caption{Polarization profile in the training data. (a) PF value distribution for 2000 uniformly sampled parameter configurations (unbalanced training data). (b) PF value distribution of 3000 training data instances, with additional 1000 parameter configurations obtained by PF value based importance sampling (balanced training data). }
% \label{PFdistribution}
% \end{figure}

\subsection{Network Structure and Training Process}
As shown in \clrr{Fig.~\ref{network_basic}(a)}, our surrogate model consists of 5 layers, comprising an input layer of 35 neurons and an output layer of 400 neurons, corresponding to the 35 simulation input parameters and 400 uniformly distributed spatial locations along the cell membrane respectively. The intermediate hidden layers, $H_0$, $H_1$ and $H_2$ are composed of 1024, 800 and 500 neurons respectively. The network is fully-connected, therefore, the neurons of one layer are connected with all the neurons of its subsequent layer. The output of each neuron in the three hidden layers are passed through ReLU (rectified linear unit) activation functions\clrr{~\cite{relu}} to model any non-linearity between the input parameters and the output responses. We also apply \textit{dropout} regularization to the first ($H_0$) and second ($H_1$) hidden layers with a dropout rate of 0.3. \clrg{Dropout regularization corresponds to randomly ignoring the activation results of the neurons in a layer during the training process to avoid over-fitted networks and achieve higher accuracy}. A dropout rate of 0.3 refers to the fact that we randomly ignore $30\%$ of the neuron outputs at layers $H_0$ and $H_1$. 

%Besides avoiding the problem of over-fitting in the training process, dropout regularization in the prediction phase can also be utilized to model the uncertainty of neural networks, which is explained in details in the next section.

We trained the surrogate model on a training dataset of size 3000. Our experts predetermined the ranges of the individual parameters and normalized them independently to the range $[-1, 1]$. The training data was created by first randomly sampling the 35 parameters in their normalized value ranges to create 3000 random parameters configurations and then running the simulation model to get the corresponding Cdc42 concentration values for each configuration ($\sim 27$ hours on a supercomputing cluster). It was observed that a large fraction of the randomly sampled training data corresponds to instances with very low Cdc42 polarization. Therefore, to train the network to predict the high polarization instances (\textbf{R1}), we performed PF value weighted training of the neural network. During the training iterations, this assigned high weightage to the loss function values corresponding to training data instances with high PF values. After training for 5000 epochs, with a batch size of 32, the model achieved a stable RMSE (root mean square error) accuracy of $87.6\%$. Standard mean squared error (MSE) was used as the loss function for training the network and the accuracy was tested on a separate validation dataset of size 500. \clrb{We tested different network architectures before finalizing on the one described in Fig.~\ref{network_basic}(a) because it had the highest accuracy and stabilized relatively faster (i.e, around $3500^{th}$ epoch). The complete accuracy profile during the training process is provided in the supplementary materials.}

\subsection{Uncertainty Quantification in Neural Network}
%The primary purpose of training the surrogate model is to create an interactive visual analysis system for the scientists to analyze and explore the yeast simulation model without having to run the expensive simulations for new parameter configurations. However, while visualizing the predicted results of the simulation model for new parameter configurations, it is important to convey the underlying uncertainty associated with the predictions. This is vital in the decision making process for the scientists, as the visual analysis system should not mislead them into believing the predicted result as the true simulation result. 

Traditionally, neural networks do not provide any measure of the uncertainty associated with its predictions by default. However, in a recent work, Gal et al.~\cite{dropout2016} showed that traditional neural networks can also be made to quantify uncertainty by activating the dropout layers in the prediction (testing) phase. As discussed in Section 5.1 above, dropout layers are generally used as regularizers in the training phase to avoid over-fitting of training data by randomly ignoring the neuron activations at different layers of the network. Dropout is generally turned off when the network makes predictions. Gal et al.~\cite{dropout2016} showed that if we apply dropout during prediction and randomly ignore the activations of neurons in different layers, we get slightly varying predictions every time the network is run with the same set of input. By observing the variations of the predicted results for multiple instances, we can quantify the uncertainty associated with the predicted results. Gal et al.~\cite{dropout2016} further proved that dropout induced uncertainty for neural networks is actually an approximation of the uncertainty obtained in Bayesian models like Gaussian Processes. Therefore, using this feature in our trained neural network-based surrogate model, we can incorporate uncertainty visualization into our proposed visual analysis system.

%Traditionally, neural networks do not provide such uncertainty information by default. As a result, surrogate models based on Gaussian Processes are more popular because they provide an inherent representation of the prediction uncertainty~\cite{}. However, in a recent work, Gal et al.~\cite{dropout2016} showed that traditional neural networks can also be made to quantify uncertainty by activating the dropout layers in the prediction (testing) phase. As discussed in Section 5.1 above, dropout layers are generally used as regularizers in the training phase to avoid over-fitting of training data by randomly ignoring the neuron activations at different layers of the network. Dropout is generally turned off when the network makes predictions. Gal et al.~\cite{dropout2016} showed that if we apply dropout during prediction and randomly ignore the activations of neurons in different layers, we get slightly varying predictions every time the network is run with the same set of input. By observing the variations of the predicted results for multiple instances, we can quantify the uncertainty associated with the predicted results. This is also referred to as Monte Carlo dropout. Gal et al.~\cite{dropout2016} further proved that dropout induced uncertainty for neural networks is actually an approximation of the uncertainty obtained in Bayesian models like Gaussian Processes. Therefore, using this feature in our trained neural network-based surrogate model, we can incorporate uncertainty visualization into our visual analysis system of the yeast simulation model.

\begin{figure}[t!]
\centering
        \includegraphics[width = 0.95\linewidth]{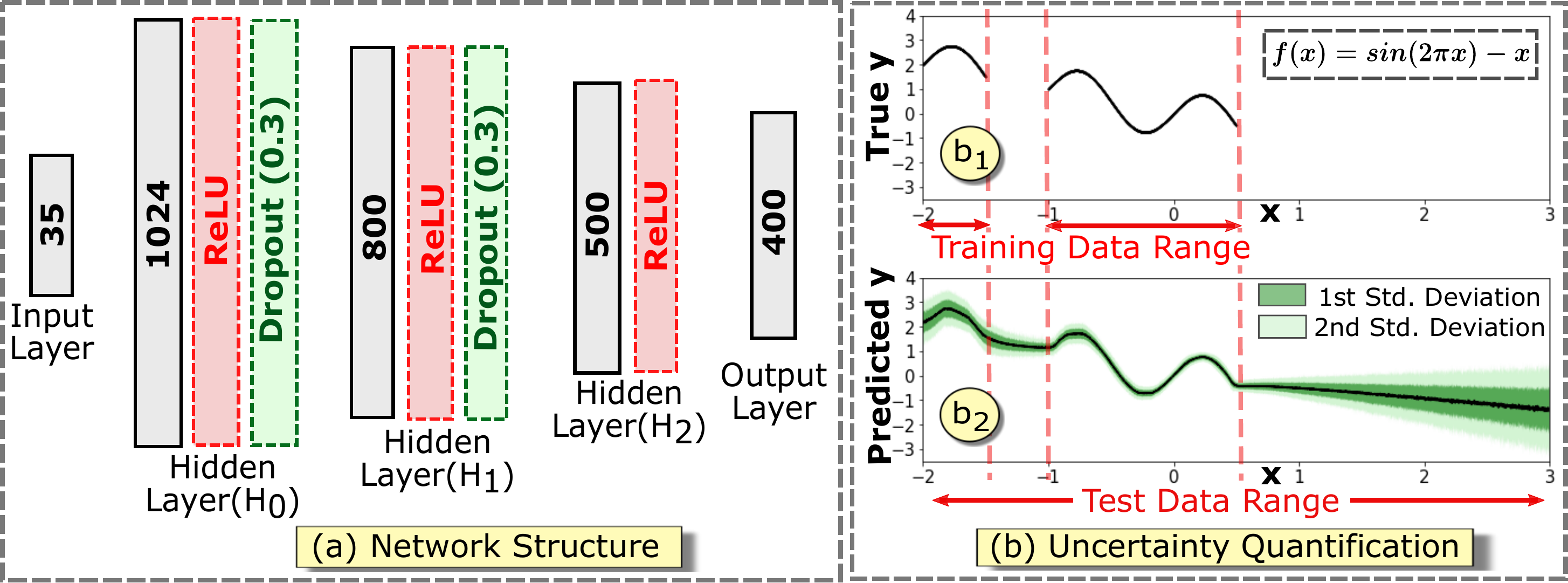} 
\caption{(a) Architecture of our surrogate model. (b) Dropout-based uncertainty visualization of neural networks for a synthetic dataset.}
\label{network_basic}
\end{figure}

% \begin{figure*}[t!]
% \centering
%         \includegraphics[width = 1\linewidth]{images/visualization/main_view_1.pdf} 
% \caption{Primary Visualizations and Interaction techniques}
% \label{visual_design_1}
% \end{figure*}

% \begin{figure*}[t!]
% \centering
% \subfloat[]{
%         \includegraphics[width = 0.47\linewidth]{images/visualization/main_view_1.pdf} }
% \subfloat[]{
%         \includegraphics[width = 0.47\linewidth]{images/visualization/main_view_weight_matrix.pdf} }
% \caption{}
% \end{figure*}

\begin{figure*}[t!]
\centering
        \includegraphics[width = 1\linewidth]{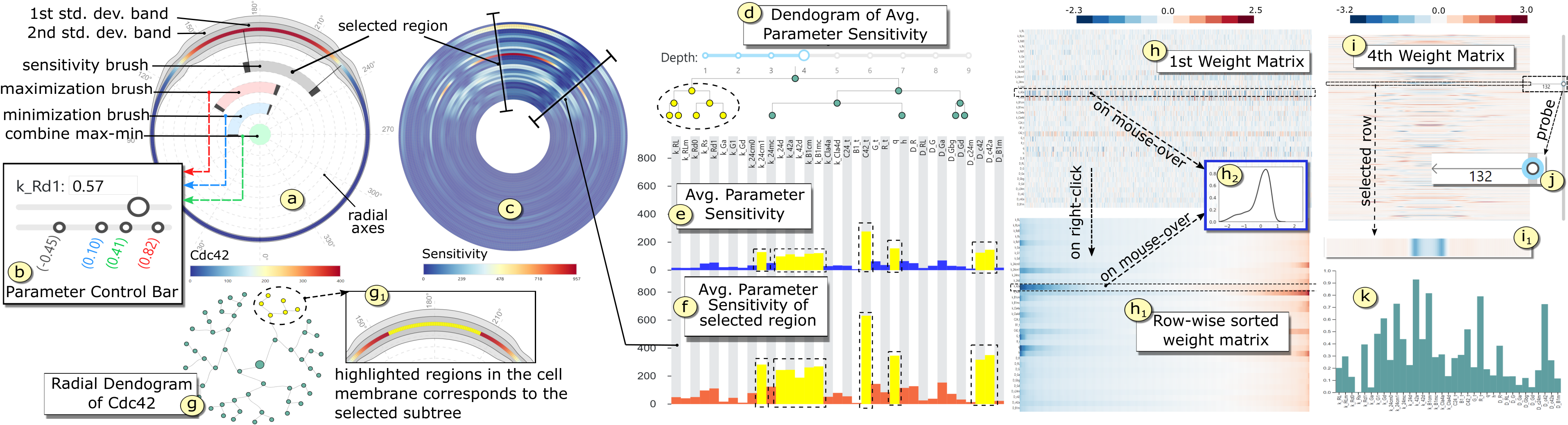} 
\caption{Primary Visualizations and Interaction techniques: (a) Predicted Cdc42 concentration across the membrane along with uncertainty bands and selection brushes. (b) Parameter control bar. (c) Spatial parameter sensitivity. (d) Linear cluster tree for average parameter sensitivity. (e,f) Average parameter sensitivities. (g) Radial cluster tree for predicted Cdc42. (h) First weight matrix. (i) Final weight matrix. (j) Row selection probe. (k) Average parameter sensitivity for selected pattern. }
\label{visual_design_1}
\end{figure*}

% \begin{figure}[t!]
% \centering
%         \includegraphics[width = 0.98\linewidth]{images/new_viz/primary_viz_vertical.pdf} 
% \caption{Primary Visualizations and Interaction techniques: (a) Predicted Cdc42 concentration across the membrane along with uncertainty bands and selection brushes. (b) Parameter control bar. (c) Spatial parameter sensitivity. (d) Linear cluster tree for average parameter sensitivity. (e,f) Average parameter sensitivities. (g) Radial cluster tree for predicted Cdc42. (h) First weight matrix. (i) Final weight matrix. (j) Row selection probe. (k) Average parameter sensitivity for selected pattern. }
% \label{}
% \end{figure}

Fig.~\ref{network_basic}(b) demonstrates the uncertainty visualization results using dropout layers in a simple 3-layer neural network on a synthetic dataset. Consider using a neural network to learn a simple sinusoidal function $f(x) = sin(2\pi x) - x$. Training data to learn this function was provided only for the $x$ ranges of $[-2.0, -1.5]$ and $[-1.0, 0.5]$. Fig.~\ref{network_basic}(b$_1$) shows the plot of the true function values $y = f(x)$ for the training data ranges. Fig.~\ref{network_basic}(b$_2$) shows the result of the trained neural network predictions for $x$ values ranging from $-2.0$ to $3.0$. The variation in the predicted result is captured using dropout layers and is visualized as standard deviation bands (in shades of green) around the mean predicted values. As can be seen, the uncertainty is high for regions where the training data was not provided to the network to learn from. The uncertainty visualization clearly shows that as we move away from the training data range (i.e, $x > 0.5$) the corresponding prediction uncertainty also increases. In order to avoid misleading the users in their decision making process and to add a sense of trustworthiness, it is vital for a visual analysis system to convey any underlying \clrr{uncertainty} in the model.

%Consider using a neural network to learn a skewed sinusoidal function $f(x)$ as shown in the top plot in figure~\ref{dropout_uncertainty}. Consider that the training data is available only between the $x$ ranges of $[-2.0, -1.5]$ and $[-1.0, 0.5]$. The top image in figure~\ref{dropout_uncertainty} shows the corresponding plot of the available training data for these example. The bottom image shows the predicted results of the neural network on a much larger value range for $x$ then the network was trained over. Dropout based modeling is used         

\subsection{Parameter Sensitivity Analysis}
\textit{Sensitivity analysis} is a popular post-hoc analysis technique performed on trained neural network models to identify the most important/salient input features. It serves as an effective tool in driving many recent advances in the field of \textit{explainable} machine learning~\cite{interpretability}. 

Sensitivity analysis of a neural network corresponds to computing the partial derivative of the outputs with respect to the inputs. Consider the $i$-th neuron in the output layer, predicting the function $f_i(\textbf{x})$ for an $n$-dimensional input vector \clrr{$\textbf{x} \sim \{x_1, ..., x_n\} \in \mathbb{R}^n$}. The sensitivity of $f_i$ with respect to the $j$-th input parameter can be denoted as $\Big(\frac{\partial f_i}{\partial x_j}\Big)^2$. A high sensitivity value corresponds to the fact that a small change in the value of the input $x_j$ is going to have a significant change in the output value of $f_i(.)$. The architecture of neural network is such that the output of every neuron in the network is completely differentiable with respect to its inputs, as a result, we can easily compute the required partial derivatives for sensitivity analysis via chain-rule using the backpropagation technique~\cite{backprop}. In our work, we utilize this to evaluate the sensitivity of the 35 simulation input parameters using the trained surrogate model. The visual analysis system provides an interface for the scientists to query for such parameter sensitivity information for different spatial regions of interest along the cell membrane. Another advantage of using neural network for sensitivity analysis is that we can also compute the sensitivity of the hidden layer activation values with respect to the input parameters as well. We utilize this to observe the parameter sensitivity towards interesting latent space data patterns learned by the hidden layers of the surrogate model during the training process.

%This measure of sensitivity is often used to identify the most important input features/parameters in a trained neural network.

\subsection{Parameter Optimization}
Another important category of post-hoc analysis operation performed on trained neural networks is called \textit{activation maximization} (AM). Activation maximization corresponds to searching for an optimal input configuration in the high-dimensional input space that maximizes the output response function. It is often used to interpret a high-level concept learned by the neural network, for example, in the field of image classification, it can be used to create new images of what the trained network thinks a cat or a dog looks like~\cite{actmax, interpretability}.      	

Neural networks are essentially optimization machines that try to find the optimal network configurations (various weight and bias values) during the training process that can best map a given input to the desired output. Once the training process is over, the network configuration is fixed and is used to predict outputs for new and unseen input configurations. Activation maximization corresponds to a reverse optimization process, where, keeping the network configuration fixed, we search for optimal input configurations that maximizes the function values of specific neurons in the output layer. For the $i$-th neuron in the output layer predicting the function $f_i(\textbf{x})$ for an $n$-dimensional input vector $\textbf{x} \in \mathbb{R}^n$, we can find an optimal input configuration $\textbf{x}^*$ by optimizing the following objective function
\begin{equation}
\max_\textbf{x} f_i(\textbf{x}) - \lambda||\textbf{x}-\textbf{x}'||^2
\end{equation}
where, the rightmost term is an $\ell_2$-norm regularizer to constrain the input search space within a known confinement $\textbf{x}'$. This penalizes the optimizer from finding an arbitrarily different $\textbf{x}^*$. The optimization involves a gradient ascent algorithm using the gradients $\frac{\partial f_i}{\partial x}$ to update the inputs to eventually find the optimal input configuration. A similar approach can also be employed to minimize the activation of a selected neuron, i.e, activation minimization, by negating the gradient values during the optimization steps. In our work, we utilize activation maximization and minimization principles to recommend simulations input parameter configurations to the scientists. They can visually selected the spatial regions in the cell membrane that they want to maximize/minimize the Cdc42 protein concentration for. By carefully choosing to maximize Cdc42 concentration in certain regions and minimize in the other regions scientists can query for parameter configurations that is likely to produce high polarization profiles in the original simulation.

%% file: tex/6_NNVA.tex
% \begin{figure}[t!]
% \centering
%         \includegraphics[width = 1\linewidth]{images/visualization/main_view_weight_matrix.pdf} 
% \caption{}
% \label{visual_design_2}
% \end{figure}

In this section, we introduce our proposed interactive visual analysis system. We first explain the primary visualizations and interaction techniques before looking at the high-level analysis views in our system.

%In this section, we introduce our proposed neural network assisted visual analysis system designed for the computational biologists to study and analyze their complex yeast cell polarization simulations. We first explain the individual visualization elements and interaction features present in our system before looking at the high-level analysis views.

%The visual design and the interaction interface are inspired by the circular shape of the cell membrane itself (figure~\ref{yeast_background}).

\subsection{Primary Visualizations and Interactions}
\textbf{\textit{Cdc42 Visualization}:} \clrb{To visualize the predicted Cdc42 concentration and preserve the circular context of the cell membrane structure, we opted for radial layout designs~\cite{radialsurvey2009}}. Fig.~\ref{visual_design_1}(a) visualizes the mean predicted concentration values for the 400 uniformly sampled points across the membrane. By default, the values are color-mapped to the maximum and minimum Cdc42 values observed in the simulation by the experts. \clrb{Radial coordinate axes (dashed-gray lines) are provided in the backdrop as frame of reference to reflect the parameterization of the simulation domain in terms of angles (degree)}. To efficiently utilize the design space, we employed the popular design principle of \textit{superposition}\clrr{, i.e.} visualizing multiple data subsets in the same coordinate system\clrr{~\cite{munzner2015visualization}}.

%To efficiently utilize the design space, we employed the popular design principles of \textit{superposition} (visualize multiple data subsets in the same coordinate system) and \textit{juxtaposition} (separate visualization of each data subset placed side-by-side).

%shows the radial layout to visualize the Cdc42 concentration predicted by the neural network based surrogate model. The mean predicted values for 400 uniformly sampled points are shown across the circumference of the circle along with radial coordinate axes (dashed-gray) in the backdrop for frame of reference.The predicted values are color-mapped to the maximum and minimum Cdc42 concentration range observed in the simulation by the scientists. 

%The example here shows an instance with good polarization profile, i.e, highly localized concentration at the top and very low in the rest of the membrane.

\textbf{\textit{Uncertainty Visualization}:} The uncertainty associated with the predicted values of the surrogate model is visualized using superimposed standard deviation bands around the circumference of the circular domain. The shapes of first (inner) and second (outer) standard deviation bands, as shown in \clrr{Fig~\ref{visual_design_1}(a)}, highlight the deviation of the predicted values in the corresponding locations of the membrane.

\textbf{\textit{Selection Brushes and Interaction}:} As shown in \clrr{Fig.~\ref{visual_design_1}(a)}, we \clrr{superimpose} multiple interactive radial selection brushes in the same coordinate system as the Cdc42 visualization. This facilitates performing various visual queries on different regions of the membrane. The \textit{sensitivity brush} (gray) allows the users to select the regions of the membrane where they want to perform \clrr{detailed} parameter sensitivity analysis. This brush is linked with the sensitivity visualizations in \clrr{Fig.~\ref{visual_design_1}(c)} and (f). The \textit{maximization brush} (red), selects the region where the users want to maximize the predicted concentration values (Section 5.4). On clicking this brush, the corresponding optimal parameter values, computed by the backend surrogate model, are reported in the parameter control bars (\clrr{Fig.~\ref{visual_design_1}(b)}). Similarly, the \textit{minimization brush} (blue) selects the region to minimize the predicted values. The green circular button at the center performs a logical AND operation of the regions selected by the maximization and minimization brushes.

%The outermost gray brush allows the scientists to select the region of interest to perform detail parameter sensitivity analysis. The \textit{sensitivity brush} is connected with the sensitivity visualization(fig~\ref{visual_design_1}(b)), which we explain in the subsequent paragraphs. The remaining brushes are related to the parameter optimization operations explained in Section 5.4. The red brush selects the region in the membrane where the scientists wants to maximize the Cdc42 concentration values, while, the blue brush corresponds to minimization of the concentration values. On left-clicking the brushes, the selection gets locked (indicated by darker shades of red and blue colors respectively) and reports the corresponding optimal parameters in the parameter control view (fig~\ref{visual_design_1}(c)). The pink circle in the center triggers an ``AND" operation on the optimization process. On left-clicking the circle (turns dark pink/magenta) it computes the optimal parameter that maximizes on the selected region by the maximization (red) brush ``AND" minimizes at the same time on the remaining regions covered by the minimization (blue) brush. The brushes need to be unlocked (by a second left-click) to make further adjustments to the selection region. Both the optimization brushes are snapped to selection sections of atleast $36^\circ$ to reduce the optimization computation overload for spatially neighboring locations which tend to generate similar optimal results.  

\textbf{\textit{Parameter Control Bars}:} Fig.~\ref{visual_design_1}(b) shows the parameter control bar for one of the simulation input parameters. The parameter name is followed by an input textbox to enter desired parameter values. The sliderbar, immediately below, can also be used to adjust the parameter values. The parameter values corresponding to different configuration instances show up in the last bar. The value corresponding to currently loaded instance show up in black colored text, whereas, the optimal parameter values recommended by activation maximization, minimization and combined max-min show up as red, blue, and green colored texts respectively. Users can click on these texts to adjust the parameter value as well. 

%The individual value ranges for the parameters were predetermined and normalized to $[-1, 1]$ by the experts. Therefore, the range of the control bars in our system is also set to the same normalized range that the experts are interested in working with.

 % The parameter name is followed by an editable input textbox to enter desired parameter values, which can also be edited using the sliderbar immediately beneath. The second bar is used to mark the parameter values for the currently loaded data (indicated by the black text color) as well as the recommended parameters by activation maximization/minimization operations. The resulting optimal parameter values from maximization, minimization and combined max-min is shown in red, blue and pink text colors respectively (corresponding to the colors of the optimization brushes discussed above). Users can click on the recommended parameter values to adjust the parameter sliderbar. As the range of all the 35 parameters have been predetermined by the scientists and pre-normalized to the range $[-1.0, 1.0]$, the value range of all the parameter sliderbars are also set to this range.     

\textbf{\textit{Sensitivity Visualizations}:} For a given parameter configuration, the local sensitivity of the 400 spatial locations across the membrane with respect to the 35 parameters are visualized in a circular heatmap as shown in \clrr{Fig.~\ref{visual_design_1}(c)}. The 35 parameters are laid out along the radial direction. A high sensitivity score for a parameter implies that a small change in its current value is going to trigger a relatively high change in the predicted Cdc42 value for the specific region of the membrane. Finer spatial selections can be made using the aforementioned \textit{sensitivity brush}. The average sensitivity of the 35 different parameters across all the spatial locations is shown as a bar-chart in \clrr{Fig.~\ref{visual_design_1}(e)}. The average parameter sensitivity information for a user selected region of interest (via sensitivity brush) is shown in \clrr{Fig.~\ref{visual_design_1}(f)}, juxtaposed with (e) to convey the relative variation in the sensitivities.

%The sensitivity selection brush in fig.~\ref{visual_design_1}(a) is directly linked with the sensitivity heatmap to highlight the selected region of interest in the membrane.  The average sensitivity of the 35 different parameters across all the spatial locations is shown as a bar-chart in fig.~\ref{visual_design_1}(e). The average parameter sensitivity information for a user selected region of interest (via sensitivity brush) is shown in fig.~\ref{visual_design_1}(f), juxtaposed with (e) to convey the relative variation in the sensitivity of the parameters.  

\textbf{\textit{Cluster Visualization}:} To help analyze the current partitioning scheme of the simulation domain (\textbf{R4}), we perform hierarchical clustering~\cite{clustering_survey} of the 400 uniformly partitioned points based on their Cdc42 values and associated uncertainty. The multi-level cluster information is visualized using a radial dendogram as shown in \clrr{Fig.~\ref{visual_design_1}(g)}. Hovering over the nodes highlights the corresponding selected clusters in the simulation domain (\clrr{Fig.~\ref{visual_design_1}(g$_1$)}). Similar clustering is performed on the average sensitivities of the 35 parameter to study their associations. Fig.~\ref{visual_design_1}(d) shows a linear dendogram view of the parameter cluster information. Users can control the tree depth, which relates to the number of clusters desired for the study. For selected nodes in the tree, the corresponding cluster members get highlighted in \clrr{Fig.~\ref{visual_design_1}(e)} and (f).

\textbf{\textit{Weight Matrix Visualization}:} To extract the knowledge learned by the trained neural network, and thereby, validate the surrogate model (\textbf{R5}), we present various techniques to analyze its weight matrices. Fig.~\ref{visual_design_1}(h) shows the weight matrix ($35 \times 1024$) between the input layer and the first hidden layer (H$_0$). The rows correspond to the 35 input parameters and the columns correspond to the final weights assigned to the 1024 neurons in H$_0$ layer. On clicking the matrix, \clrb{the weights are sorted in ascending order for each row of the matrix}, which helps identify interesting weight distribution patterns as shown in \clrr{Fig.~\ref{visual_design_1}(h$_1$)}. \clrb{Detailed explanation about the interesting weight patterns and their importance is provided in Section 7.2}. On hovering the mouse over the rows, we display the shape of the corresponding weight distribution for the respective parameter in a pop-up window (\clrr{Fig.~\ref{visual_design_1}(h$_2$)}). Based on the patterns observed in the final weight matrix ($500 \times 400$) between the last hidden layer (H$_2$) and the output layer, as shown in \clrr{Fig.~\ref{visual_design_1}(i)}, we offer a separate set of interactions to study the matrix. We provide a row selection probe in the form of a sliderbar with an arrowhead (\clrr{Fig.~\ref{visual_design_1}(j)}) to select the rows with interesting weight patterns. The selected row index is highlighted in the side and a zoomed-in view of the row is displayed (\clrr{Fig.~\ref{visual_design_1}(i$_1$)}). Fig.~\ref{visual_design_1}(k) shows the average parameter sensitivity chart corresponding to the neuron in the penultimate layer (H$_2$), which is responsible for the selected weight pattern.

\textbf{\textit{Colormap Adjustments}:} We offer two sets of value ranges for mappings the colors when visualizing the predicted Cdc42 concentration. The first set (default) is based on the minimum and maximum concentration values that the experts feel is effective for studying cell polarization behavior. For the current system, this range is set to $[0,400]$. The second set corresponds to the local minimum and maximum values for individual prediction results. This provides more control to the experts during the analysis period, because the concentration values for different instances can have varying dynamic ranges. \clrb{We have included nine divergent and three sequential colormaps~\cite{cmap} in our system for the users to choose from.}

%We also offer nine different carefully selected divergent colormap~\cite{cmap} alternatives as well as three  for the Cdc42 visualization (\clrr{Fig.~\ref{visual_design_1}a}) and the spatial sensitivity heatmap (\clrr{Fig.~\ref{visual_design_1}c}). 

\subsection{Visual Analysis System} 
Using the visualizations and interaction techniques explained above, we design our visual analysis system with multiple high-level views to provide a structured and efficient analysis workflow for our experts. It comprises of the following high-level views, each addressing different facets of the analysis requirements set forth in Section 4.

\begin{figure}[t!]
\centering
        \includegraphics[width = 0.95\linewidth]{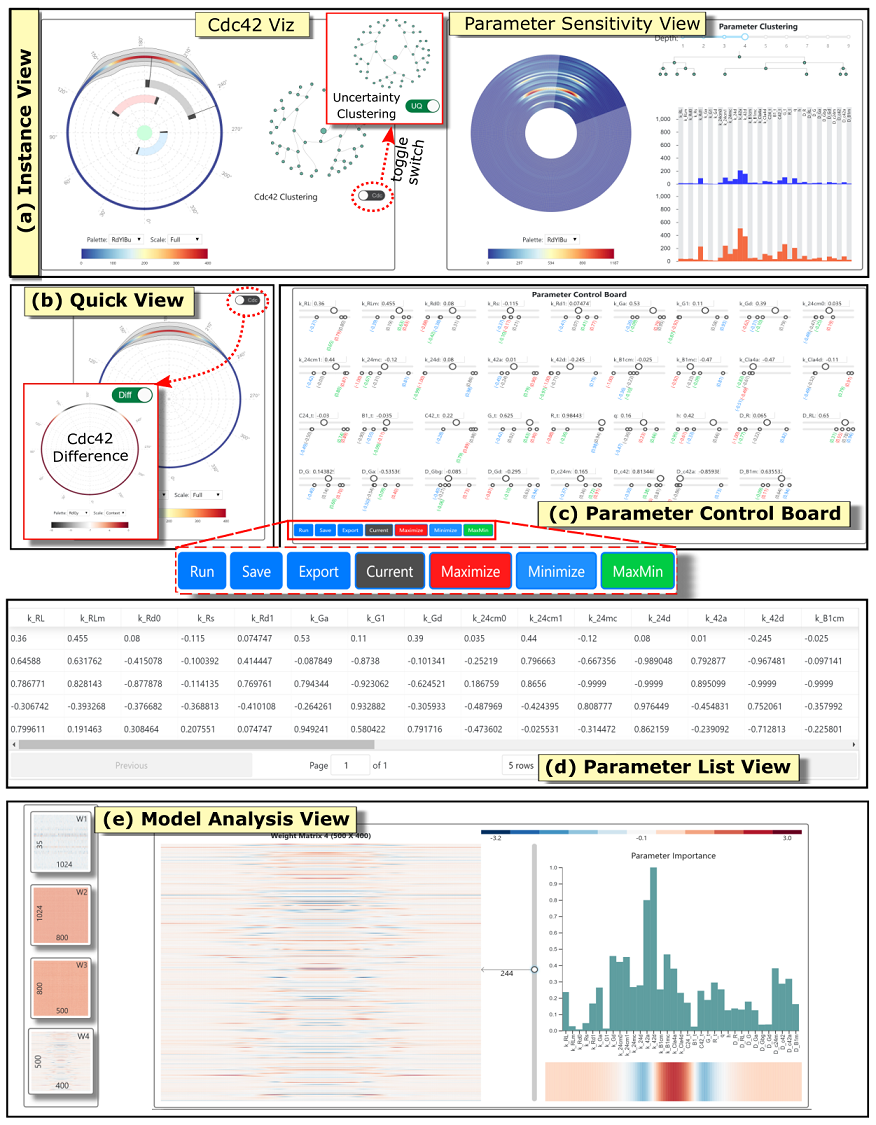} 
\caption{Multiple high-level analysis views of our visual analysis system.}
\label{NNVA}
\end{figure}

\textbf{\textit{Instance View}:} This view loads the visualizations corresponding to a specific \textit{instance} of input parameter configuration that the scientists wish to analyze in detail (\textbf{R3}, \textbf{R4}). As shown in \clrr{Fig.~\ref{NNVA}(a)}, it comprises of two sub-views. \textit{Cdc42 Viz} displays the predicted Cdc42 concentration result for the currently analyzed parameter instance, whereas, \textit{Parameter Sensitivity View} displays the corresponding spatial sensitivity as well as the overall average parameter sensitivity for the specific instance. Both the sub-views have the corresponding cluster trees described in Section 6.1 for detail analysis. In \textit{Cdc42 Viz}, there is a switch to toggle between the radial cluster trees corresponding to the Cdc42 values and the uncertainty values (\textbf{R4}).

\begin{figure*}[t!]
\centering
        \includegraphics[width = 0.98\linewidth]{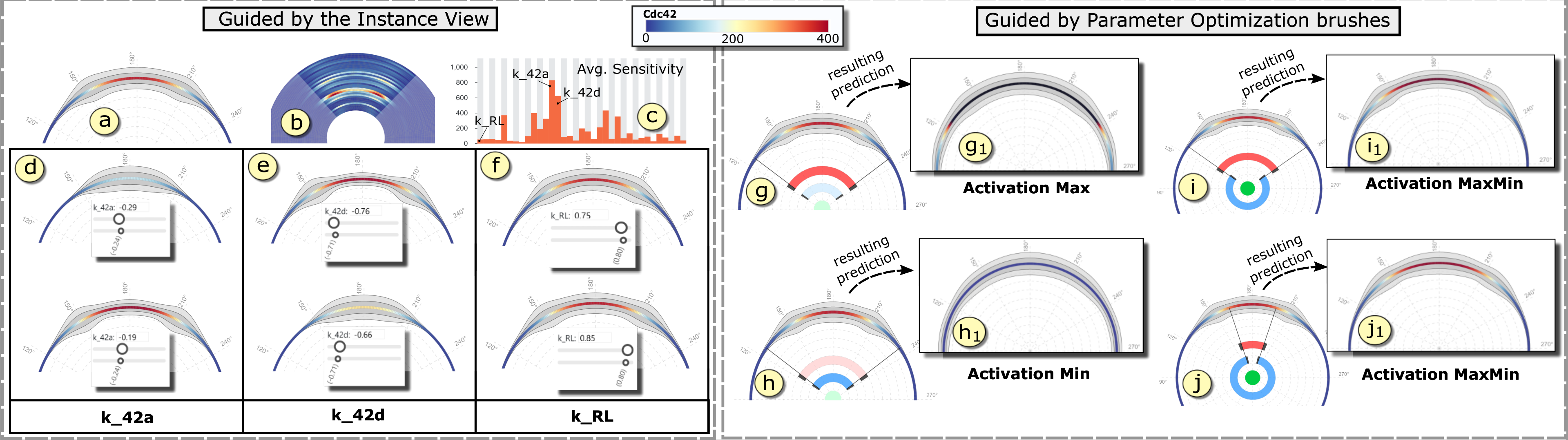} 
\caption{Discover new parameter configurations: (a) Predicted Cdc42 of a specific parameter instance with relatively high polarization profile. (b) Spatial parameter sensitivity of the parameter instance. (c) Corresponding average parameter sensitivities. Results for slightly changing the highly sensitive parameters \texttt{k\_42a}(d), \texttt{k\_42d}(e) and a less sensitive parameter \texttt{k\_RL}(f). Maximizing (g) and minimizing (h) predicted Cdc42 values in the selected regions. (i,j) Maximizing and minimizing the predicted values for the selected regions \textit{at the same time} to get highly polarized predictions (i$_1$, j$_1$).  }
\label{cs1}
\end{figure*}    

%\textbf{Instance View:} This view offers in-depth analysis and visualization of the predicted Cdc42 concentration for a specific \textit{instance} of the input parameter configuration that the scientists want to study in details (R3). As shown in Fig.~\ref{NNVA}a, the instance view comprises of two sub-views. ``Cdc42 Viz" displays the predicted protein concentration across the cell membrane along with uncertainty, while, the ``Parameter Sensitivity View" shows the detailed spatial sensitivity of the specific input parameter configuration. The corresponding parameter configuration instance shows up in the ``Parameter Control Board" view discussed below. Both the sub-views have detail clustering informations in the form of dendograms. For ``Cdc42 Viz", there are two versions of the radial dendogram corresponding to the Cdc42 concentration values and the associated uncertainty values across the membrane respectively. Users can toggle between these two types of tree-views using the switch provided on the lower right-side of the radial tree. 

%``Cdc42 Viz" also comprises of all the selection brushes discussed in Section 6.1. The ``Parameter Sensitivity View" consists of the sensitivity visualizations and the linear dendogram structure of parameter sensitivity clustering as discussed in Section 6.1.

\textbf{\textit{Parameter Control Board}:} This view serves as the main panel to visualize and interactively modify the input parameter configurations (\textbf{R1}). As shown in \clrr{Fig.~\ref{NNVA}(c)}, the 35 different \textit{parameter control bars} (Section 6.1) for the individual parameters are laid out across four rows. It visualizes the parameter values corresponding to the parameter instance currently analyzed in the \textit{Instance View} as well as the optimal parameters recommended by the interactive optimization brushes. Users can modify the parameter values and click the \textit{Run} button to execute the neural network-based surrogate model in the backend to generate the corresponding simulation prediction, which is visualized in the \textit{Quick View} (\textbf{R2}). The \textit{Save} button lets us store the modified input configuration in the \textit{Parameter List View}, whereas, the \textit{Export} button downloads the list of saved configurations. Instead of manually adjusting the 35 different parameter sliders, users can click the \textit{Current}, \textit{Maximize}, \textit{Minimize} or \textit{MaxMin} buttons to automatically set the parameter sliders/values to the desired recommended configurations.

\textbf{\textit{Quick View}:} As shown in \clrr{Fig.~\ref{NNVA}(b)}, it visualizes the predicted Cdc42 concentration for the user-modified parameter configurations via the \textit{Parameter Control Board}. This offers a means to perform rapid prototyping of the expensive simulation using the surrogate model, thus facilitating exploratory analysis with new and unseen parameter configurations (\textbf{R1}, \textbf{R2}). In order to compare the predicted Cdc42 values vis-\`a-vis the results in the \textit{Instance View}, we provide a toggle switch to change the visualization in \textit{Quick View} to display the exact difference in Cdc42 concentration across the membrane.

%\textbf{Quick View:} As mentioned above, ``Quick View" (fig.~\ref{NNVA}b) displays the predicted Cdc42 concentration results for the user-modified parameter configurations via the "Parameter Control Board". This offers a means to perform rapid prototyping of the expensive simulation results using the surrogate model, thus facilitating exploratory analysis of the simulation model for new and unseen parameter configurations (R1, R2). A toggle switch is provided at the top right corner to switch between the predicted Cdc42 visualization and the visualization of the difference in Cdc42 values vis-a-vis the specific instance analyzed in the ``Instance View" above. Visualizing the difference in Cdc42 values with respect to the results of a known parameter instance allows the scientists to infer the prediction quality of the modified parameter configure, which is essential to discover new parameter setting for the simulation (R1).   

\textbf{\textit{Parameter List View}:} As shown in \clrr{Fig.~\ref{NNVA}(d)}, this view temporarily stores the newly discovered parameter configurations. Additionally, users can click on the rows in the list to load the selected configuration back in the \textit{Parameter Control Board}. This list of configurations can be exported/downloaded in a file format which can be directly used to execute the original simulation model. This offers a seamless integration between the analysis workflow involving our visual analysis system and the actual simulation modeling workflow of the experts (\textbf{R1}).

%\textbf{Parameter List View:} The discovered parameter configurations can be temporarily stored in the ``Parameter List View" using the \bluebutton{Save} button in ``Parameter Control Board". Fig.~\ref{NNVA}d shows the stored list of parameter configurations. Additionally, users can click on the rows in the list to load the selected configuration back in the ``Parameter Control Board".  This list of parameter configurations can be exported as configuration files which can be directly fed to the original yeast simulation model for execution. This offers a seamless analysis workflow for the scientists to discover new parameter configurations (R1) using our visual analysis system and integrating with their simulation modeling activities. 

\textbf{\textit{Model Analysis View}:} This view lets the users investigate the trained neural network-based surrogate model to extract useful insights about the simulation model (\textbf{R5}). As shown in \clrr{Fig.~\ref{NNVA}(e)}, the left-most panel shows the thumbnail views of all the weight matrices of the trained network. Users can click on the matrix images to open up the corresponding analysis views in the right panel. Detailed analysis, as explained in Section 6.1, can be performed on the selected weight matrix to extract any data patterns and validate the knowledge learned by the surrogate model.

%It also provides our experts a way to validate the surrogate model behavior by evaluating the knowledge learned by the model against their domain knowledge. 

%\textbf{Model Analysis View:} Fig.~\ref{NNVA}e shows the surrogate model analysis view, where the users can diagnose the trained neural network to extract useful information learned by the model during the training process. The left-most panel shows the thumbnail views of all the weight matrices of the trained netwo. As can be seen, the first and the last matrices show distinct patterns in their weight distributions compared to the second and third matrices. Users can click on the matrix images to open up the corresponding analysis views in the right panel. Detail analysis, as explained in Section 6.1, can be performed on the selected weight matrix to extract any data patterns and validate the knowledge learned by the surrogate model.

%% file: tex/7_case_study.tex
In this section, we perform two case studies using our visual analysis system and evaluate the results by comparing against the original simulation outcomes as well as the findings from a previous polynomial surrogate model based analysis of the simulation~\cite{renardy2018}. 
%In this section, we perform two case studies using our visual analysis system to identify new simulation input parameter configurations and analyze the black-box of the neural network to extract interesting properties of the simulation model. We also perform extensive evaluation of the results in both the case studies by comparing against the results generated by the true simulation model as well as the results of a previous analysis effort using polynomial surrogate model~\cite{renardy2018}.  

\subsection{Discover New Parameter Configurations}
One of the key requirements from the system is to visually guide the users towards discovering desired parameter configurations (\textbf{R1}), instead of having to perform random sampling of the high-dimensional parameter space. This also enables the experts to incorporate their domain knowledge into the parameter discovery process. In this case study, we use our system to identify new parameter configurations that can trigger high Cdc42 polarization in the original simulation model. We support two different visual parameter discovery approaches. According to the conceptual framework of Sedlmair et al.~\cite{sedlmair2014visual}, both of these approaches can be categorized under \textit{local-to-global} and \textit{informed trial and error} visual parameter navigation strategies.

\textbf{Guided by the Instance View:} In this approach, users can center their parameter discovery process around a known instance of input parameter configuration, whose results get loaded in the \textit{Instance View}. The \textit{Instance View} offers in-depth analysis of the Cdc42 concentration as well as the corresponding parameter sensitivity information for the loaded parameter instance. Fig.~\ref{cs1}(a), (b) and (c) show the zoomed-in views of the predicted protein concentration, spatial parameter sensitivity and average parameter sensitivity respectively of a selected region for the loaded parameter instance. \clrb{As can be seen in Fig.~\ref{cs1}(a)}, the loaded instance corresponds to a good polarization profile. Using this as the starting point, users can start modifying the parameter configurations based on the parameter sensitivity details of the loaded instance combined with their domain knowledge. 

For example, from the sensitivity views in \clrr{Fig.~\ref{cs1}(b)} and (c), we can infer that the parameters \texttt{k\_42a} and \texttt{k\_42d} are the most sensitive parameters for the loaded configuration. This implies that a small change to these parameter values will significantly change the current polarization profile. We use the \textit{Quick View} panel to visualize the predicted polarization profile generated by the modified parameter values. Fig.~\ref{cs1}(d) shows the results for decreasing and increasing the current \texttt{k\_42a} value ($-0.24$) by a step-size of $0.05$ in the top and bottom images respectively. Decreasing the parameter value significantly brings down the protein concentration at the top of the cell, while, increasing the parameter value increases the concentration. A reverse trend is observed for \texttt{k\_42d} (Fig.~\ref{cs1}(e)), where increasing the parameter values by $0.05$ reduces the concentration and vice-versa. 

From the simulation perspective, this behavior makes sense because \texttt{k\_42a} corresponds to Cdc42 activation, whereas, \texttt{k\_42d} corresponds to Cdc42 deactivation. Since the type of simulated protein is active Cdc42 (i.e, C42a), \texttt{k\_42a} have a positive impact on its concentration while \texttt{k\_42d} have a negative impact. Similarly, for a less sensitive parameter like \texttt{k\_RL}, we can see in \clrr{Fig.~\ref{cs1}(f)} that changing its parameter value by the same step-size does not result in a significant change in the polarization profile. Using this approach, we identified 8 new parameter configurations which are likely to produce high Cdc42 polarization results in the original simulation.

%As discussed in Section 6.2, the toggle switch in the \textit{Quick View} panel helps visualize the exact difference in the concentration values with respect to the \textit{Instance View}, thereby, conveying the relative change in concentration.

\begin{figure}[t!]
\centering
        \includegraphics[width = 0.95\linewidth]{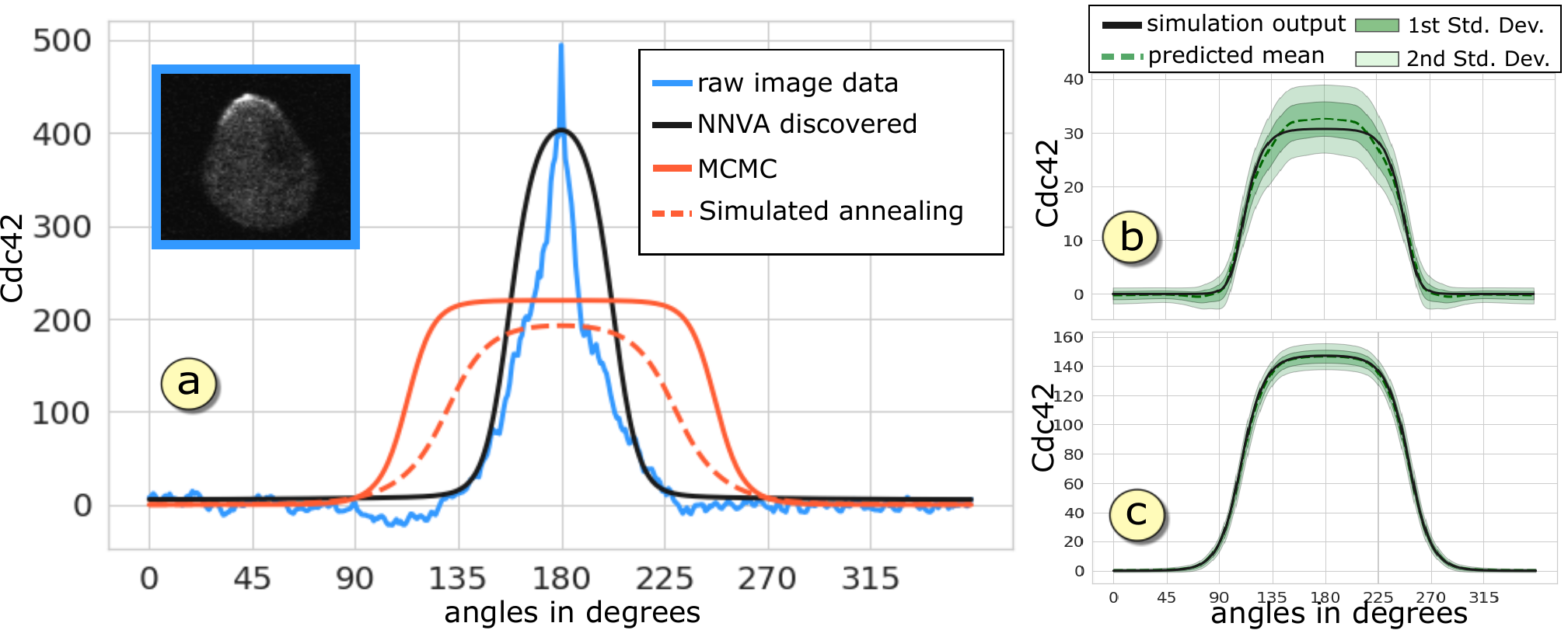} 
\caption{(a) Comparative evaluation of the simulation results using parameter configurations discovered by our system (black) and previous analysis work (red)~\cite{renardy2018}. Comparison curves of Cdc42 concentration for (b) a highly uncertain prediction and (c) a good prediction instance. }
\label{cs1_eval}
\end{figure}

\textbf{Guided by Parameter Optimization Brushes:}  We also recommend new parameter configurations based on the activation maximization/minimization analysis framework of neural network described in Section 5.4. Users can utilize the maximization brush, as shown in \clrr{Fig.~\ref{cs1}(g)}, to select the region of the cell membrane where they want to maximize the predicted Cdc42 value. The corresponding optimal parameter configurations computed by the surrogate is recommended in the \textit{Parameter Control Board}. \clrg{Fig.~\ref{cs1}(g$_1$) shows the result of running the surrogate model with the recommended parameters.} \clrb{As can be seen in Fig.~\ref{cs1}(g$_1$)}, the predicted Cdc42 values in the selected region sharply increases (even beyond the maximum value of 400, set by the experts). Similarly, a minimization brush, as shown in \clrr{Fig.~\ref{cs1}(h)}, recommends a parameter configuration that brings down the concentration values of the selected region close to the minimum concentration value (\clrr{Fig.~\ref{cs1}(h$_1$)}). For the particular case of finding high polarization profiles, we are interested in maximizing the concentration at the top of the cell and minimizing the concentration across rest of the locations at the same time. We make the desired selection using the two brushes to make this query as shown in \clrr{Fig.~\ref{cs1}(i)} and (j) for two different selection ranges. The corresponding recommended parameters display high Cdc42 polarization predictions, as indicated by the \textit{Quick View} results in \clrr{Fig.~\ref{cs1}(i$_1$)} and (j$_1$) respectively for the two different selections. On top of the recommended parameters, users can further modify individual parameter values to create new sets of parameter configurations. Using this approach, we created a total of 7 new input parameter configurations.

\textbf{Evaluation:} We executed the simulation model using the 15 newly identified parameter configurations and observed high degrees of Cdc42 polarization (PF $> 0.5$) for all the configurations. We found 5 parameter configurations with PF values exceeding 0.8, whereas, in the initial training data that was collected by randomly sampling the parameter space, we never found an instance with PF value close to 0.8. The highest PF value recorded among the newly discovered parameters was 0.82 and corresponds to the optimized configuration recommended by the neural network for the selection shown in \clrr{Fig.~\ref{cs1}(j)}. In \clrr{Fig.~\ref{cs1_eval}(a)}, we compare the actual simulation result generated with our discovered optimal parameter configuration versus using the parameters estimated by previous polynomial surrogate model based analysis~\cite{renardy2018}. The sharp blue curve (PF = 0.87) corresponds to the protein concentration obtained by extracting the pixel intensity values of a real microscopic image of a highly polarized yeast cell as shown in the blue box. This acts as the ground truth for the simulation model to generate similar levels of polarization results. The black plot shows the simulation result produced using the optimal parameters discovered by our system (PF = 0.82). The solid red plot (PF = 0.57) and dashed red plot (PF = 0.64) corresponds to the simulation results generated by the parameter configurations estimated in previous work~\cite{renardy2018}. 

This is a significant improvement over the previous parameter analysis results for the same simulation model. This establishes that, given the right input parameter setting, the simulation model is capable of generating sharp polarization results similar to real laboratory results. We also visually verified the predicted results of the surrogate model against the original simulation outputs by plotting them together as curves. Fig.~\ref{cs1_eval}(b) and (c) show the comparative plots for a highly uncertain prediction and a less uncertain prediction instance respectively. The green dashed-line shows the mean prediction curve, whereas, the solid black line is the original simulation output.

 %Fig.~\ref{cs1_eval}(a) compares multiple plots of the Cdc42 concentration produced by the simulation model, where the x-axis corresponds to the angles along the computational domain and the y-axis is the actual Cdc42 values. The noisy blue plot corresponds to the pixel intensity values extracted from a real microscopic image data of a highly polarized cell as shown in the blue box.  The PF values of this curve as computed by equation~\ref{eq1} is 0.87. This acts as the ground truth for the simulation model to generate similar levels of polarization results. The black plot shows the simulation result generated by the parameter configuration discovered using our neural network assisted visual analysis system (PF = 0.82). While solid and dashed red plots (with PF values 0.57 and 0.64 respectively) where the simulation results with the parameters estimated using a polynomial surrogate model analysis. This is a significant improvement over the previous parameter analysis results for the same simulation model, thus proving that the simulation model with the newly recommended parameters is able to simulated high polarization results similar to laboratory results. 

\subsection{Knowledge Extraction from Surrogate Model}
In this case study, we aim to extract the knowledge learned by the network during its training process by analyzing its various weight matrices. This also serves the purpose of validating the surrogate model to make sure that it is making predictions based on some reasonable domain-aligned logic rather than random ad-hoc predictions (\textbf{R5}).

%The trained neural network not only learns to predict the simulation output for new parameters, but also learns interesting data patterns and parameter properties during its training process. In this case study, we aim to extract such knowledge by analyzing its various weight matrices. This also serves the purpose of validating the surrogate model to make sure that it is making predictions based on some reasonable domain-aligned logic rather than random ad-hoc predictions (\textbf{R5}).

\begin{figure}[t!]
\centering
        \includegraphics[width = 0.85\linewidth]{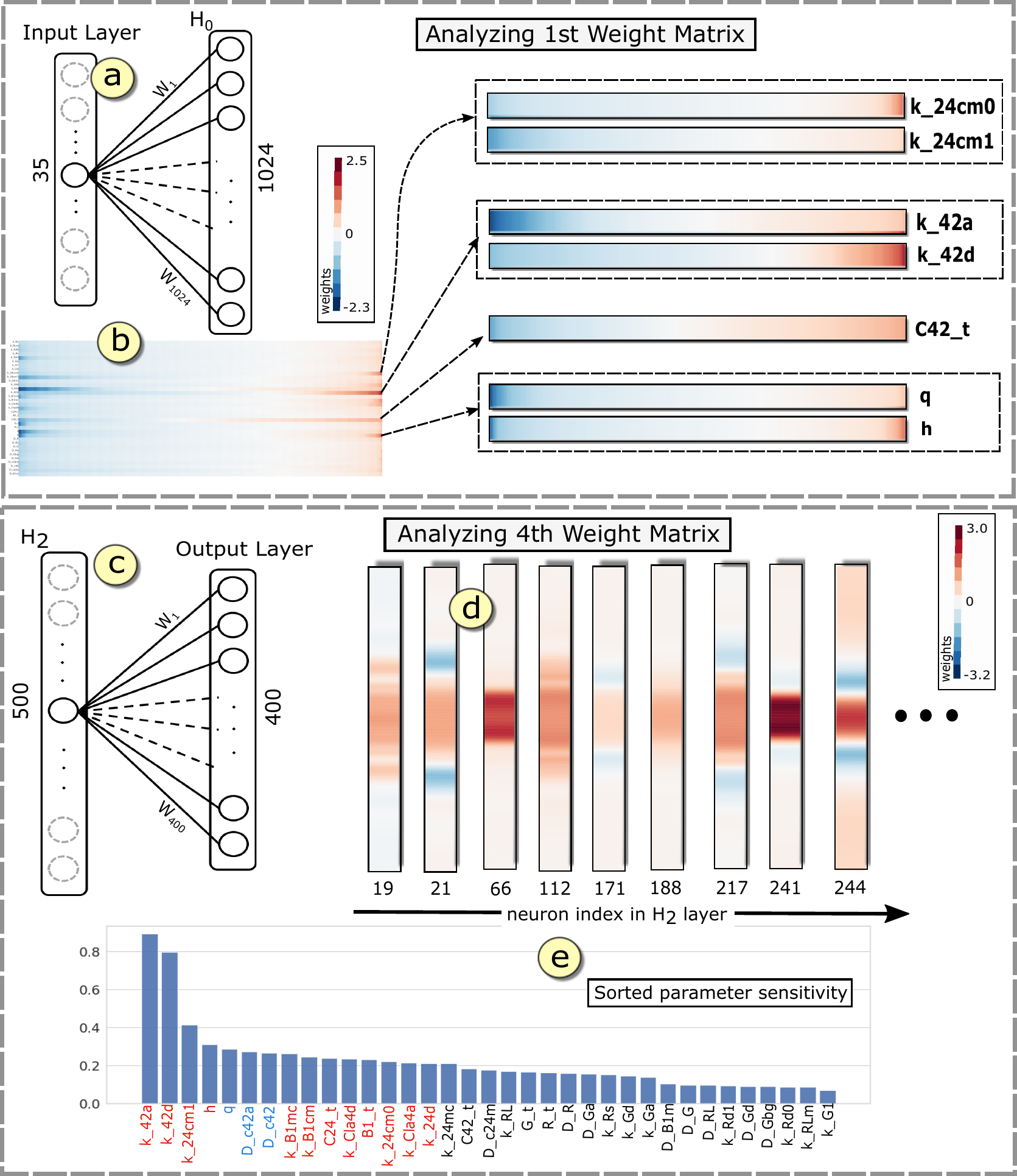} 
\caption{Knowledge extraction: (a) Connections of one parameter with H$_0$ layer. (b) Row-wise sorted first weight matrix. (c) Connections of a neuron in H$_2$ layer with the output layer. (d) Few selected weight patterns with high weights at the center. (e) Corresponding average parameter sensitivity sorted in descending order.}
\label{cs2}
\end{figure}

\textbf{First Weight Matrix:} In the first weight matrix between the input layer and first hidden layer (\clrr{Fig.~\ref{cs2}(a)}), we observe distinct patterns in the distribution of weights for certain parameters. Fig.~\ref{cs2}(b) shows the full $35 \times 1024$ matrix with the weight values sorted in ascending order for each row (i.e, parameter) to highlight the patterns (\clrr{Fig.~\ref{visual_design_1}(h)} shows the original matrix view). We observed relatively high negative and positive weights for parameters \texttt{k\_24cm0}, \texttt{k\_24cm1}, \texttt{k\_42a}, \texttt{k\_42d}, \texttt{C42\_t}, \texttt{q} and \texttt{h}. High positive and negative weights in the first matrix for a parameter corresponds to the fact that the parameter values had to be scaled by the weights for some neurons in the first hidden layers. This implies that the original range of values provided for the parameters is not sufficient. Similar observations were made for these parameters explicitly in the previous analysis work~\cite{renardy2018}. The experts feel that the range of these parameters need to be expanded to get better simulation results. Besides, the matrix also verified that the pairs (\texttt{k\_24cm0}, \texttt{k\_24cm1}), (\texttt{k\_42a}, \texttt{k\_42d}), and (\texttt{q}, \texttt{h}) showed relatively strong correlation compared to the other parameters. The exact shape of the weight distributions for these parameters are provided in the supplementary materials.

\textbf{Final Weight Matrix:} We did not observe any distinct patterns in the second and the third weight matrices. However, in the final weight matrix of resolution $500 \times 400$, corresponding to the third hidden layer and the output layer, we found multiple interesting weight distribution patterns (full matrix view is shown in \clrr{Fig.~\ref{visual_design_1}(i)}). Each row of this matrix corresponds to the weights assigned to individual neuron activation values in the penultimate layer ($H_2$) towards the 400 output neurons, as illustrated in \clrr{Fig.~\ref{cs2}(c)}. We observed that different neurons in $H_2$ layer assign high positive or high negative weights to different sets of output neurons. In the output layer, the neurons at the middle section corresponds to the top of the cell membrane modeled by the simulation. Therefore, it is interesting to identify which neurons in the $H_2$ layer assigns high positive weights to the middle section of the output layer, because those neurons are most likely to contribute towards producing high Cdc42 polarization results. We identified 95 such neurons in the $H_2$ layer (\clrr{Fig.~\ref{cs2}(d)}) and evaluated their average parameter sensitivity (Section 5.3) to find out which parameters are more sensitive to produce the selected weight pattern in the penultimate layer. 

Fig.~\ref{cs2}(e) shows the sorted list of parameters based on descending order of their average normalized sensitivities. We compared this importance/sensitivity order of parameters for generating high polarization patterns with that of the list of highly sensitive parameters identified in the previous work of our experts~\cite{renardy2018}. We found that except the change in order of the 3 parameters \texttt{q}, \texttt{D\_c42a}, and \texttt{D\_c42} by one position (marked by blue texts), the top 15 sensitive parameters (red texts) were in the same order as previously identified by experts. Similar analysis can be perform for different weight distribution patterns in the penultimate layer of the network. 

%Besides analyzing the global sensitivity of parameters for high polarization patterns, we can utilize our visual analysis framework to study many other interesting weight patterns, for example, instances with high negative weights assigned to the top of the cell membrane. This flexibility enables the experts to not only utilize the system to study Cdc42 behavior, but also other protein species in the future, which can show different polarization patterns. 

%% file: tex/8_feedback.tex
\clrb{Our experts from the field of computational biology comprise of a professor from the Department of Mathematics, who created the yeast simulation model, and two of her graduate students.} They feel that the proposed visual analysis system is a very useful tool to fine-tune their simulation model. \clrb{They found the visual interface of our system to be simple and intuitive for users familiar with yeast simulation models.} The ability to quickly prototype different parameter combinations and interactively visualize the predicted simulation output within seconds lets them easily calibrate the simulation model. 

Previously, there was no interactive visualization created for the yeast simulation. Our experts feel that the visual analysis system will be a useful medium to communicate with the non-expert collaborators and stakeholders of the project, instead of explaining them the complex reaction-diffusion equations involved in the simulation. As discussed in Section 7.1, using our system we were able to discover new parameter configurations that can trigger high Cdc42 polarization in the original simulation model. This is a significant improvement over the previously estimated parameters using polynomial surrogate model analysis~\cite{renardy2018}.

The experts feel that the system is flexible to work with other protein species besides Cdc42. The current backend, i.e, neural network-based surrogate model, predicting Cdc42 concentration, can be easily replaced with another neural network model predicting different species and still retain the same visual interactions to analyze the simulation. Since the backend analysis techniques are independent of the network structure, they can train a network with a different architecture and still utilize our visual analysis frontend. The experts also plan to utilize the radial clustering information to create an adaptive mesh for the computational domain rather than the current 400 uniformly spaced resolution.

The model analysis view was helpful to validate the trained network and see if the surrogate model is actually learning something relevant about the simulation rather than making random predictions. However, they feel that the model analysis view requires users to have a good understanding of the neural network architecture to interpret the weight matrices. They feel it would be helpful to make the weight matrix analysis more intuitive for people without much machine learning background. Overall, the experts are satisfied with our neural network assisted visual analysis system and feel that it meets all of their expected requirements. They plan to use the findings from the visual analysis system to improve the simulation and report in a systems biology journal in future.   

%% file: tex/9_discussion.tex
\begin{figure}[t!]
\centering
        \includegraphics[width = 1\linewidth]{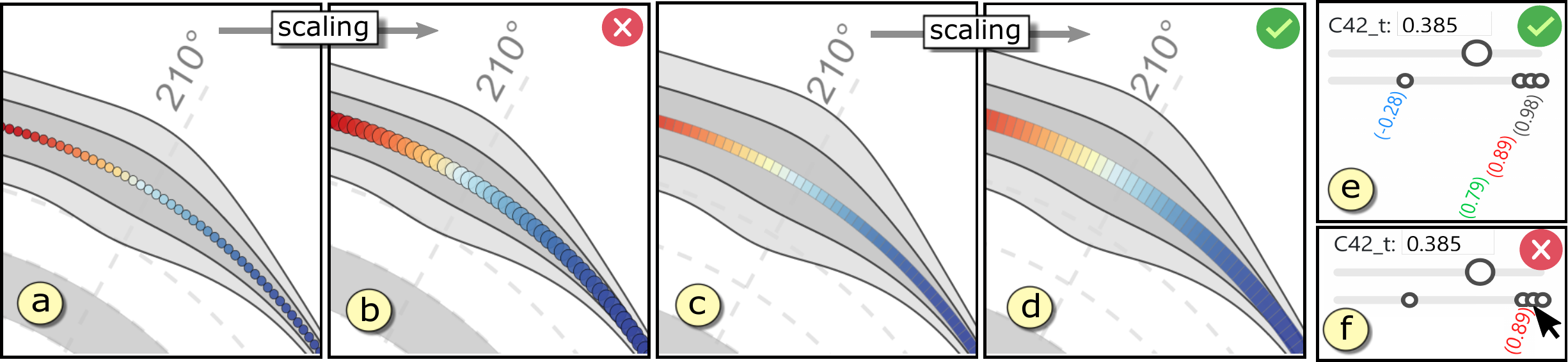} 
\caption{Design study: Using circles (a,b) versus using rectangular boxes (c,d) across the membrane. Parameter control bar with (e) all the values displayed versus using (f) mouse-hovering.  }
\label{design_study}
\end{figure}

\quad \clrb{\textbf{Design Choices:} The choice of using radial layouts for some of the visualizations is inspired from the circular shape of the computational domain (Fig.~\ref{yeast_background}(c)). This helps retain the context of the cellular structure during the visual analysis workflow. Throughout the course of this project, we iterated over different design choices for various elements of our visual analysis system. Following are some of the key design decisions that we had to make in the process.}
\begin{itemize}
	\setlength\itemsep{0.05em}
	\item \clrb{Initially, we used 400 colored circles along the circumference of the simulation domain to visualize the concentration values (Fig.~\ref{design_study}(a)). However, as the overall system grew in size, we had to reduce the size of the individual views, which significantly shrunk the size of the 400 small circles. Increasing the size of the small circles led to overlapping among the spatially neighboring points (Fig.~\ref{design_study}(b)). Therefore, we changed the visual design to use contiguous rectangular boxes instead of circles at each point (Fig.~\ref{design_study}(c)). As a result, we can scale the boxes radially without worrying about overlapping (Fig.~\ref{design_study}(d)). }

	\item \clrb{Deciding on an optimal design layout for the \textit{Parameter Control Board} with 35 different \text{parameter control bars} was a challenging task. One straight-forward choice was to place the control bars for 35 parameters in a long vertical/horizontal panel with scrollbars to scroll through the list of individual control bars. However, the experts felt that it is important to have all the 35 bars visible at the same time, without the need to scroll around during the model calibration process.}

	\item \clrb{The next challenge was to find the best way of showing multiple recommended parameter values in the same view. Fig.~\ref{design_study}(e) shows an instance in our current system with 3 recommended parameter values very close to each other. We lay them out vertically to make the close-by values standout. This helps the users in clicking on the texts and adjusting the parameter slider to that precise recommended value. Another choice to save space in the \textit{Parameter Control Board} panel was to show the values only when the mouse hovers over the recommended nodes as shown in Fig.~\ref{design_study}(f). However, we went with the first choice as it helps the experts to see all the recommended parameter values in the same view and not just the nodes in the bar. We believe that there could be better designs to address this, and we plan to explore other alternatives in the next version of our system.}
\end{itemize}

\clrb{\textbf{Comparison with Previous Works:} Previous visual analysis systems for simulation parameter exploration were all specifically designed to meet the requirements of their respective domain applications. However, one popular choice among some of the systems~\cite{orban2019drag,piringer2010hypermoval,berger2011uncertainty} was to project the high-dimensional space to a low-dimensional space for visualization. We did not opt for this choice because our experts were interested in directly manipulating in the exact parameter space rather than in some latent space. This helps them to explicitly map the effect of the parameters to different output regions. Using the reduced latent space can be confusing to interpret the meaning in the high-dimensional space. One unique feature of our proposed approach is that all the \textit{analysis tasks} and \textit{navigation strategies}~\cite{sedlmair2014visual} supported in our system are carried out using a single analysis framework in the backend, i.e, the trained neural network. Whereas, when using other surrogate models for prediction, most of these backend activities have to be carried out separately~\cite{biswas2018interactive,berger2011uncertainty,biswas2017visualization,sedlmair2014visual}. In this work, we analyzed a simulation with 35 input parameters. To the best of our knowledge, previous visual parameter analysis systems~\cite{sedlmair2014visual,orban2019drag,biswas2018interactive,wang2017multi} did not have to deal with such large number of simulation parameters.}

%One main advantage of our proposed neural network-assisted visual analysis system is that, all of these analysis and navigation tasks are carried out using a single analysis framework in the backend, i.e, the trained neural network-based surrogate model.

\textbf{Implementation and Performance:} We used the \textit{Keras} python library (\textit{v2.2.4})~\cite{keras}, with \textit{TensorFlow} backend, to implement and train our neural network-based surrogate model. We trained the model on a NVIDIA Pascal P100 GPU for 5000 epochs. It took 59.31 minutes to train the network for 5000 epochs and the final accuracy of the model was $87.6\%$. To perform post-hoc operations like sensitivity analysis and activation maximization on the trained network, we used the \textit{Keras-Vis}~\cite{kerasvis}, which is a high-level analysis library for neural networks. To perform uncertainty quantification for neural networks, we wrote custom code to turn on the dropout layers for trained \textit{Keras} models during prediction/testing phase. Our frontend visual analysis system was designed using \textit{d3.js}. We used \textit{flask} framework~\cite{flask} to interact with the trained neural network from our visual analysis system. The average time to get the predicted simulation output from the trained neural network for a new parameter configuration is 0.25 seconds, whereas, running the original simulation model for one configuration took 2.3 hours in a supercomputing cluster.

%The full accuracy plot during the training process is provided in the supplementary materials.

%% file: tex/10_conclusion.tex
In this paper, we have proposed an interactive visual analysis system to study and analyze a complex yeast cell polarization simulation model. The proposed system uses a trained neural network-based surrogate model as the backend analysis framework to facilitate interactive visual analysis. It allows the experts to interactively calibrate the simulation input parameters as well visually guide them towards discovering new parameter configurations. We also analyze the surrogate model to extract interesting insights about the original simulation model. We hope that our proposed approach can motivate researchers to look at neural networks as more than a prediction tool, and start utilizing them to conduct interesting analysis activities.

%We applied various post-hoc analysis techniques like uncertainty quantification, sensitivity analysis, and activation maximization for trained neural networks to drive our visual analysis system.

In future, we would like to extend the visual analysis framework to facilitate more complex analysis tasks like the recently proposed \textit{testing with concept activation vectors} (TCAV)~\cite{tcav}. This will allow the experts to validate high-level domain specific concepts in the surrogate model. We plan to apply similar visual analytic approach for analyzing simulations from other application domains as well. As suggested by our experts, we also plan to simplify the model analysis and validation methods so that it is more intuitive for people without much machine learning background.  